# Robust characterization of photonic integrated circuits

Jiajia Wang,[1,#] Xingyuan Xu,[1,#,*] Haoran Zhang,[1] Xuecheng Zeng,[1] Yunping Bai,[1] Arthur J. Lowery,[2,*] Kun Xu,[1,*]

[1]State Key Laboratory of Information Photonics and Optical Communications, Beijing University of Posts and Telecommunications, Beijing 100876, China

[2]Electro-Photonics Laboratory, Department of Electrical and Computer Systems Engineering, Monash University, Clayton, 3800 VIC, Australia

[#]These authors contribute equally to this work.

*Corresponding author: xingyuanxu@bupt.edu.cn, arthur.lowery@monash.edu, xukun@bupt.edu.cn

**Keywords:** programmable photonic integrated circuits, phase recovery, robust characterization

**Abstract:**
Photonic integrated circuits (PICs) offer ultra-broad optical bandwidths that enable unprecedented data throughputs for signal processing applications. Dynamic reconfigurability enables compensation of fabrication flaws and fluctuating external environments, tuning for adaptive equalization and training of optical neural networks. The initial step in PIC reconfiguration entails measuring its dynamic performance, often described by its frequency response. While measuring the amplitude response is straightforward, e.g. using a tunable laser and optical power meter, measuring the phase response presents challenges due to various factors, including phase variations in test connections and instrumentation limitations. To address these challenges, a universal and robust characterization technique is proposed, which uses an on-chip reference path coupled to the signal processing core (SPC), with a delay larger or smaller than the total delay across the signal processing paths. A Fourier transform of the chip's power response reveals the SPC's impulse response. The method is more robust against low reference-path power and imprecise delays. Experiments using a finite-impulse-response (FIR) structure demonstrate rapid SPC training, overcoming thermal crosstalk and device imperfections. This approach offers a promising solution for PIC characterization, facilitating expedited physical parameter training for advanced applications in communications and optical neural networks.

## Introduction

Photonic integrated circuits (PICs), [1-5] which harness the speed, bandwidth, and low energy consumption of optical systems, offer a paradigm shift in signal processing for applications ranging from telecommunications to artificial intelligence.[6-20] Among the many desired attributes of PICs, dynamic reconfigurability stands as a fundamental requirement for practical applications. This necessity arises to, for example, (a) compensate for imperfections within PICs, such as fabrication inaccuracies and thermal crosstalk; (b) adapt to temperature fluctuations; (c) update synaptic weights for neural network hardware. [21-25]

The primary step in reconfiguring a PIC involves determining its key functional parameters, perhaps using indirect measurements. [26] For example, the weights of each optical path in a finite impulse response (FIR) filter can be determined using a Fourier transform of its frequency response, provided that this response contains both amplitude and phase information. While the amplitude response of a PIC can be straightforwardly obtained with a wavelength-swept laser and photodiode, capturing the phase response poses challenges due to phase variations in measurement paths (such as coupling fibers), as well as inaccuracies (optics) or limited bandwidths (electronics) of external measurement instruments. Thus, retrieving the phase response from the amplitude response, rather than measuring it directly, can significantly mitigate complexity during the PIC's characterization process.

Recent approaches for PIC characterization have used on-chip reference paths but have limitations: (a) the Kramers-Kronig (K-K) phase retrieval method [27] requires a strong signal in the reference path to sustain the Minimum Phase (MP) condition, resulting in weak optical power allocation to the signal processing core (SPC) and consequently degraded phase recovery accuracies in noise-limited scenarios, thereby limiting the scale of the PICs; (b) the fractional-delay reference path imposes a stringent requirement on the PIC's delays, [28] thus is non-universal—so only applicable to PICs with uniformly spaced discrete delays.

Here we propose a universal and robust characterization technique for complex PICs. This is achieved through an on-chip reference path coupled in parallel with a signal processing core (SPC). The impulse response of the SPC can be obtained by directly performing a Fourier transform of the entire chip's power response, provided the reference path is sufficiently short or long. This approach alleviates the stringent requirements on the PIC's power and delays inherent in prior approaches, offering key advantages: (a) significantly increased power fed to the SPC, enhancing robustness against noise; (b) universality and compatibility with PICs featuring non-integer spaced delays. The feasibility and robustness of our approach were validated using a finite-impulse-response (FIR) structure as the SPC; leveraging the characterized information and feedback algorithms, the SPC was trained to execute distinct signal processing functions within approximately 50 iterations, surmounting severe thermal crosstalk and device imperfections. Our approach represents an optimal characterization for PICs, presenting robust and universal pathways towards characterizing and expedited training of PICs, which is crucial for advanced applications in communications and optical neural networks.

**Theory: optimized characterizing method for programmable PICs**

As shown in Fig. 1, the overall on-chip system consists of a signal processing core (SPC) coupled in parallel with a reference path via a tunable coupler. Two pairs of on-chip ports can be used: one pair accessing the whole chip for characterization and thus calibration, and the other pair accessing solely the SPC for signal processing.

The reference path carries a power equal to that of the SPC (taking SPC's loss into account) and a tailored delay to introduce a temporal "gap" in the impulse response (thus this technique is named as "gap method"). The gap has a duration $\tau$ between the reference tap and SPC, which is larger than the SPC's temporal duration $T$. Considering that the SPC's loss is generally much higher than the reference waveguide, the majority of input optical power during calibration is fed into the SPC.

The SPC has a frequency response of $H_{spc}$ and can be implemented in diverse forms, such as finite impulse response (FIR) filters, unitary transformers and waveguide meshes.[29] The power response $|H_{chip}|^2$ of the whole chip (i.e., the insertion loss spectrum that is straightforward to measure via a wavelength-swept laser and a power meter) can be regarded as the superposition of frequency-domain raised-sinusoidal responses originating from: "internal" interferences between pairs of SPC taps; and "external" interferences between the reference tap and each of the SPC taps.

By performing a Fourier transform (FT) of the measured power frequency response $|H_{chip}|^2$, a temporal series or impulse response can be obtained, which contains the impulse response information of the SPC corresponds to the "external" interferences and interfered spurious terms corresponds to the "internal" interferences. Due to the designed delay gap ($\tau > T$), the spurious terms fall within the "gap" between the reference and SPC paths' impulses, so can be discarded. As such, the needed information—exact phases and amplitudes (i.e., complex tap coefficients) of the SPC—can be obtained. Importantly, a complex-valued inverse FT is used unconventionally; that is, its real-valued optical-field input fed from the power response (i.e. field-squared) measurement, and its imaginary input is set to zero. This is entirely valid because all resulting distortions fall within the gap. A similar technique has been used to implement optical OFDM (Orthogonal Frequency Division Multiplexed) transmission systems using only intensity detection, though the time and frequency domains are reversed in that case. [30]

Assisted by a windowing technique (see Supplementary Materials), the gap method does not require stringent alignment/correlation between $\tau$ and $T$, thus does not impose any restrictions on the SPC's timing sequence. We note that, for the power response measurement, the frequency resolution $\Delta f$ needs to satisfy $1/\Delta f > 2(T+\tau)$, as required by the Nyquist sampling theorem.

Practical application scenarios are generally noise-limited, where the external measurement system has a certain noise floor, and the PICs have non-negligible losses, such as PICs with many cascaded on-chip components to form splitting and combining trees. To evaluate the gap method's performance in noise-limited scenarios, we denote the power difference between the taps and the noise of the impulse response as the "Temporal

Signal-to-Noise Ratio" (TSNR, measured by OVA, assuming the power of noise and the total power of taps is constant, the *TSNR* of the whole chip is constant). Specifically, the optimal performance of SPC characterization can be obtained via our gap method, as the power of the reference path and the SPC are equal (see Supplementary Materials for detailed proofs).

We note that the Kramers-Kronig (K-K) method [31] is frequently invalid in this optimum operation regime, where the Minimum Phase condition $|h_{ref}| > |H_{spc}|$ can no longer be met, especially in sampled-data systems. [32] For example, for an FIR sinc filter with $N$ taps, the Minimum Phase condition indicates that the reference path's amplitude needs to be higher than $N$ times of the SPC's amplitude: assume the amplitude of each SPC's tap is $E$, then $|H_{spc}|=N \cdot E$, $|H_{spc}|^2 = N^2 \cdot E^2$; in the worst case of all paths having equal phase (which could occur at some particular frequency), the amplitude of reference path needed to support the MP condition should satisfy $|h_{ref}| > |H_{spc}| = N \cdot E$, thus its power $|h_{ref}|^2 > N^2 \cdot E^2$, which is $N^2$ times of SPC's power. This will deteriorate the accuracy of the characterization in noise-limited scenarios, especially when considering filters with many taps (i.e., large $N$), required to support finely-detailed frequency responses such as sharp-transition filters.

**Experiments**

Fig. 2a shows the topology of the PIC. The delay step of the FIR filter's taps (i.e., the minimum delay step of the SPC), $\Delta T$, is 20.4 ps. This gives the SPC's impulse response duration $T = 142.8$ ps. The gap is roughly set as $\tau \approx 1.13 \times T$ to illustrate the tolerance of our approach in terms of the reference path's delay. Four stages of tunable couplers were employed to control the power allocation between the reference path and delay paths and eight thermal phase shifters were employed to adjust the SPC taps' phases.

For the insertion loss measurement, the frequency resolution $\Delta f$ was set as 160 MHz; thus, $1/\Delta f = 6.25$ ns $> 2(T+\tau) = 0.61$ ns, satisfying the Nyquist sampling criterion; the measurement range was set as 170.9 GHz $\approx 3.5/\Delta T$.

We set the amplitudes of all SPC taps to be equal. The electrical power applied onto the first-stage tunable MZI (highlighted in red in Fig. 2a) was swept with a range larger than $P_{2\pi}$ (the power needed to achieve $2\pi$ phase shifts for one arm of the MZI) to change the power splitting ratio between the reference path and SPC.

Fig. 2c shows the power of impulse response as calculated by an optical vector network analyzer, from its frequency-response measurement. The SPC paths have equal power, as desired. The temporal gap between the reference and the SPC impulses is clearly visible. The *TSNR* of the whole chip was calculated to be 24.98 dB. At the optimal operating point with equal power distributions, the power of the reference path and the SPC were equal.

Fig. 2d shows the maximum values of the chip's zeros (grey line, obtained via Z-transform of the recovered impulse responses) with varying power splitting ratios between the reference path and the SPC. These reveal the minimum powers needed to sustain the minimum phase condition (i.e., all zeros need to be within the unitary circle).

In the region where the K-K method remains valid (blue shaded, Fig. 2d-e), the reference path's power is >8.3 dB higher than that of the SPC (>10×$\log_{10}N$ = 9 dB for ideal sinc filters). However, the optimal operation points for the gap method, where the power of the reference path and SPC are equal, are in the region where the K-K method is no longer valid (red shaded, Fig. 2d-e).

The root-mean-square error (RMSE) of the recovered taps' amplitudes is plotted with measured results by the optical vector network analyzer in Fig. 2e. We note that, as the error performance of both methods is subject to the power splitting ratio between the reference path and the SPC, for regions with significantly wide power distributions, the power difference between the reference path and SPC (>10 dB, slash shaded, Fig. 2e), will be more likely to be limited by noise.

Fig. 2f shows measured insertion loss spectra of the whole chip (measured via the calibration ports of the chip), corresponding impulse responses (recovered via the gap and K-K methods) and pole-zero plots at five specific electrical power values/power splitting ratios labeled in Fig. 2d. Plots (*ii*) and (*iv*) are cases at the optimal operation points (denoted by the RMSE results in Fig. 2e)—accessible only by the gap method: Plots (*i*) and (*v*) are cases where the reference path is much stronger in power than the SPC, and thus both methods are valid; Plot (*iii*) is the case where the reference path is >20 dB weaker in power than SPC such that both methods' recovered impulse responses are buried in noise.

The results clearly show: (a) the gap method's performance is always better than the K-K method; (b) the best error performance is obtained by the gap method at the optimal operation points, where the K-K method is invalid because the reference is too weak to meet the minimum phase condition.

We performed simulations to further verify the above claims. For parameter values similar to the experiments (tap coefficients, MZI power splitting ratio vs heater power, and *TSNR*), the simulated SPC has characteristics, including the MP threshold (Fig. 3a) and the RMSE of SPC's tap amplitudes (Fig. 3b), similar to the experimental results, again verifying the superiority of the gap method. We further varied *TSNR* at various reference-SPC power splitting ratios (operation points (*i*) and (*ii*)) to compare the performance of the two SPC characterization methods. As shown in Fig. 3d-e, the calculated RMSE of tap coefficients clearly shows that: (a) the gap method's performance is always better than the K-K method; (b) the gap method is more accurate at poor *TSNR*.

Part of the value of self-calibration, as provided by this and previous methods, [27] is that it allows a chip's characteristics to be accurately set to a desired value or "dialed-up". [33] With the on-chip information accurately recovered via the gap method, we trained the SPC to achieve diverse (arbitrary, in theory) transfer functions for signal processing, including sinc, multi-passband, low-pass, bandpass filters, and a Hilbert transformer. During each training iteration, the retrieved and desired impulse responses of the SPC were first mapped as practical and desired parameters of on-chip tunable devices (i.e., power splitting ratio of MZIs and phase shifts of phase shifters), respectively; then the required updates of electrical power can be inferred and applied onto the heaters. By iterative measuring and updating the SPC's dynamic parameters, server on-chip thermal crosstalk effects and device fabrication inaccuracies can be compensated.

As shown in Fig. 4 (first row for amplitude, second row for phase), the tap coefficients of the FIR signal processing core converged to the desired values within 50 training iterations, at which point the distinct amplitude and phase distribution of the taps can be observed. The relatively fast convergence of the self-learning process is a result of the successful recovery of the chip's tap coefficients, which enables direct calculation of the errors between the practical system (taken dynamic thermal crosstalk and fabrication errors into account) and desired functions and thus the needed updating amounts of electrical power applied onto the chip.

After the training process, the SPC tap coefficients (Fig. 4, fourth and fifth rows), retrieved from the whole chip's amplitude response (Fig. 4, third row), converged to desired values; and thus, the SPC was dialed up to achieve target frequency responses (Fig. 4, sixth and seventh rows).

## Conclusion

We have proposed a robust and universal characterization technique for PICs and demonstrated the approach with a FIR structure coupled in parallel with an integrated reference path. The impulse response of the FIR structure can be accurately retrieved via the Fourier transform of the chip's measured power response. Server on-chip thermal crosstalk was overcome via iterative calibration algorithms, enabling diverse signal processing functions being accurately implemented. This work paves a road towards on-line training of PICs for demanding applications in neuromorphic computing, quantum computing and optical communications.

## Methods

The PIC was fabricated in CUMEC (Chongqing United Microelectronics Center) using a standard Silicon-On-Insulator platform, with a total chip footprint of 25 mm$^2$. TiN heaters (each with ~1.7 kΩ resistance) are used to tune the optical phases via thermo-optical effects. The fabricated PIC is a 16-tap FIR chip. For a proof-of-concept demonstration using a standard chip design, we used the shortest path as the reference path and only half of the taps as the SPC. The unused paths were suppressed in power and should be removed in future implementations. Luceda IPKISS was used to lay out the chip.

During the experiments, the chip was mounted on an external thermal controller, with the temperature stabilized at 30 °C to dissipate the heat generated by the on-chip phase shifters. The optical insertion loss spectra (i.e., the power response or the squared amplitude response) of the chip were measured by an optical vector network analyzer (Luna OVA 5100), which can be replaced with a wavelength-swept laser and an optical power meter, as we introduced above.

The *TSNR* was calculated from the impulse responses (power) captured by Luna OVA as the ratio between temporal power densities of signal (i.e., taps) and noise. The temporal power density of noise was denoted by the mean power of the noise floor; the temporal power density of the signal was calculated in several steps: (i) integrate the impulse responses (power) in the range that the taps exist; (ii) divide the value obtained in (i) by the length of the impulse responses; (iii) substrate the value obtained in (ii) by the temporal power density of noise.

## Supporting Information

Supporting Information is available from the Wiley Online Library or from the author.

**Acknowledgements:** This work was supported by National Key R&D Program of China (No. 2021YFF0901700), National Natural Science Foundation of China (No. 62301074, 61821001,62135009), Fund of State Key Laboratory of IPOC (BUPT) (No. IPOC2023ZZ01).

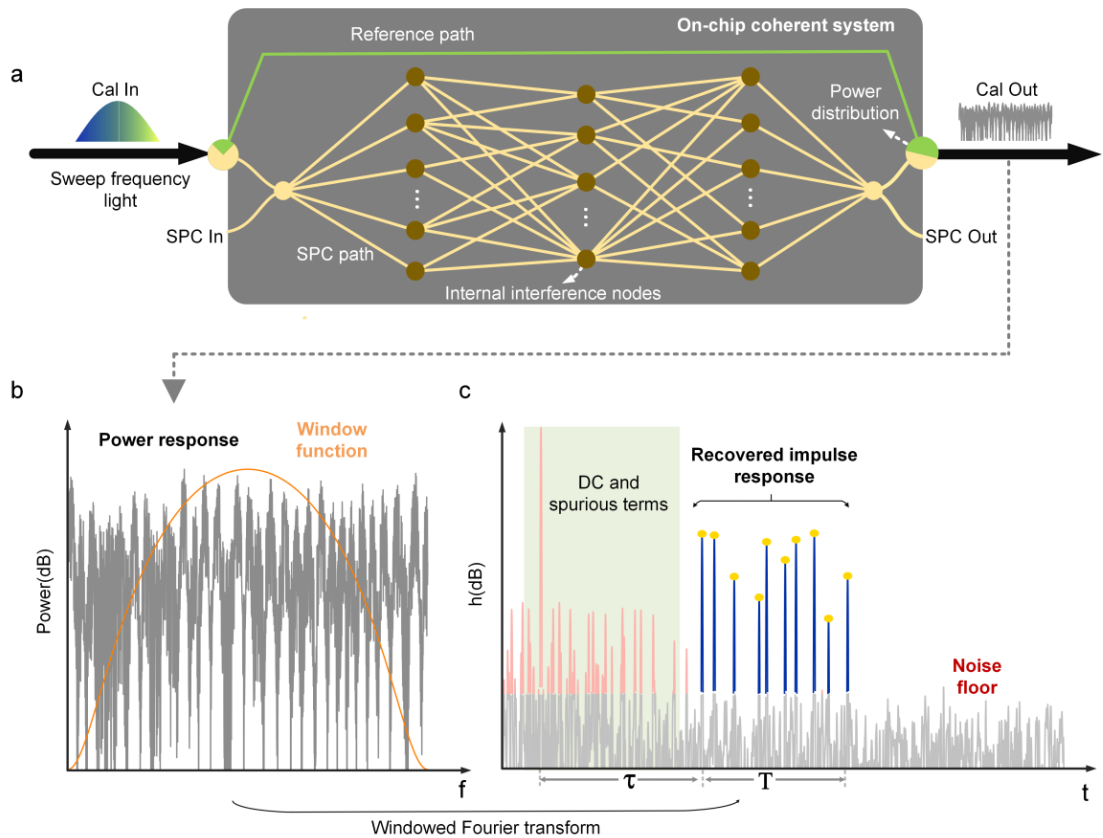

**Figure 1| PIC characterization process.** (a) Conceptual signal flow. The system comprises an on-chip reference path and the signal processing core (SPC). An on-chip tunable coupler enables the power allocation between the reference path and the SPC. (b)   Power response versus frequency. (c) Fourier transform of the power response (noting that the power response is sent into the real-field input of the Fourier transform). For a long/short-enough reference path, the desired impulse response of the SPC (blue) is distinct from the spurious terms (red).

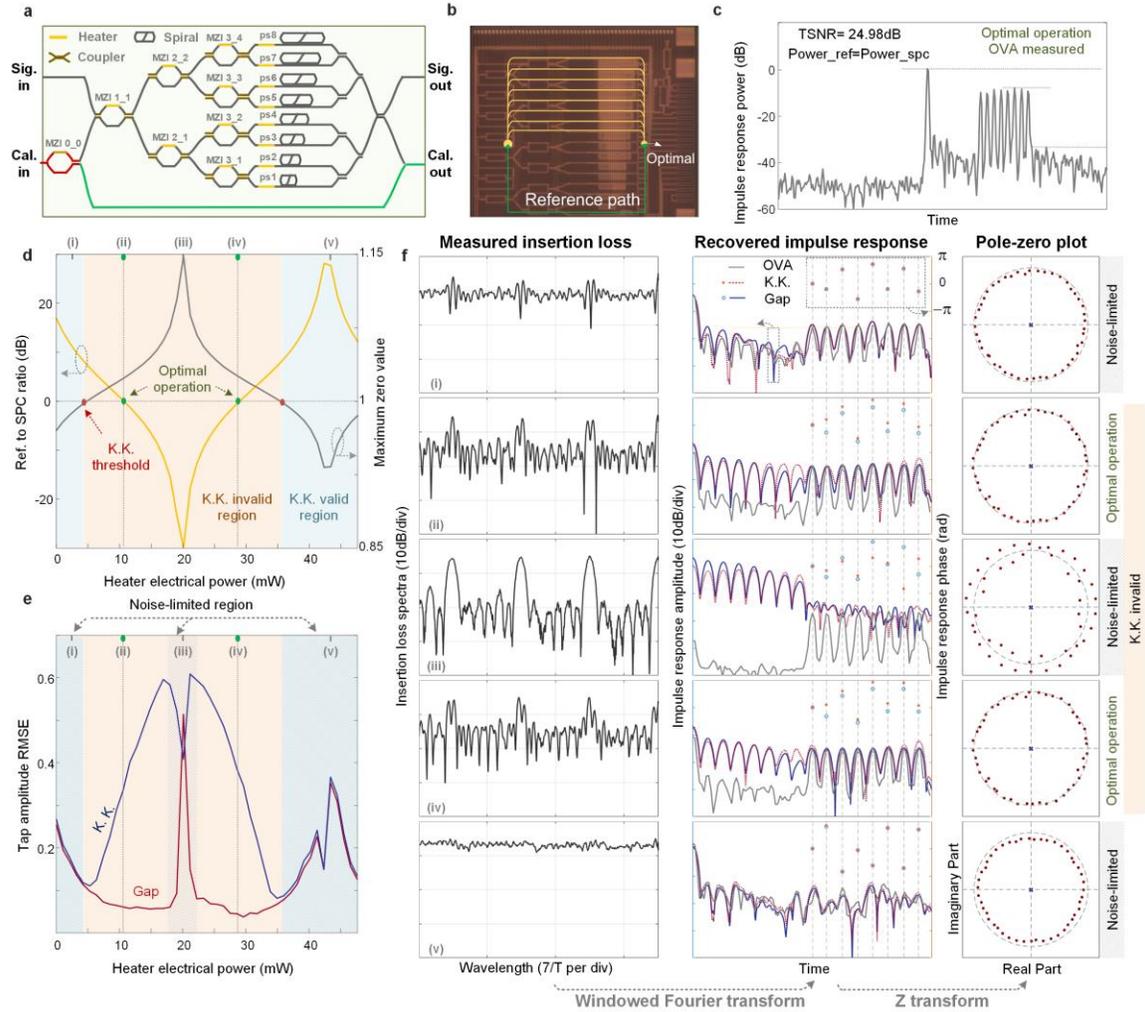

**Figure 2 | Experimental results of PIC characterization using gap and K-K methods.** (a) Schematic of the PIC's layout. (b) Optical micrograph of the PIC. (c) Impulse response power measured by the optical vector network analyzer, with the chip operating at the optimal power distributions. (d) The measured power splitting ratio between the reference path and SPC (yellow line) and the maximum value of zeros (grey line). (e) The calculated RMSE of retrieved tap amplitudes using the K-K method (blue line) and our gap method (red line). (f) The measured insertion loss spectral (left column), retrieved impulse responses (middle column), and pole-zero plot (right column). Note that due to the non-integer delay of the gap, we padded zeros to make the system equivalent to one with uniform delays, for the pole-zero plot.

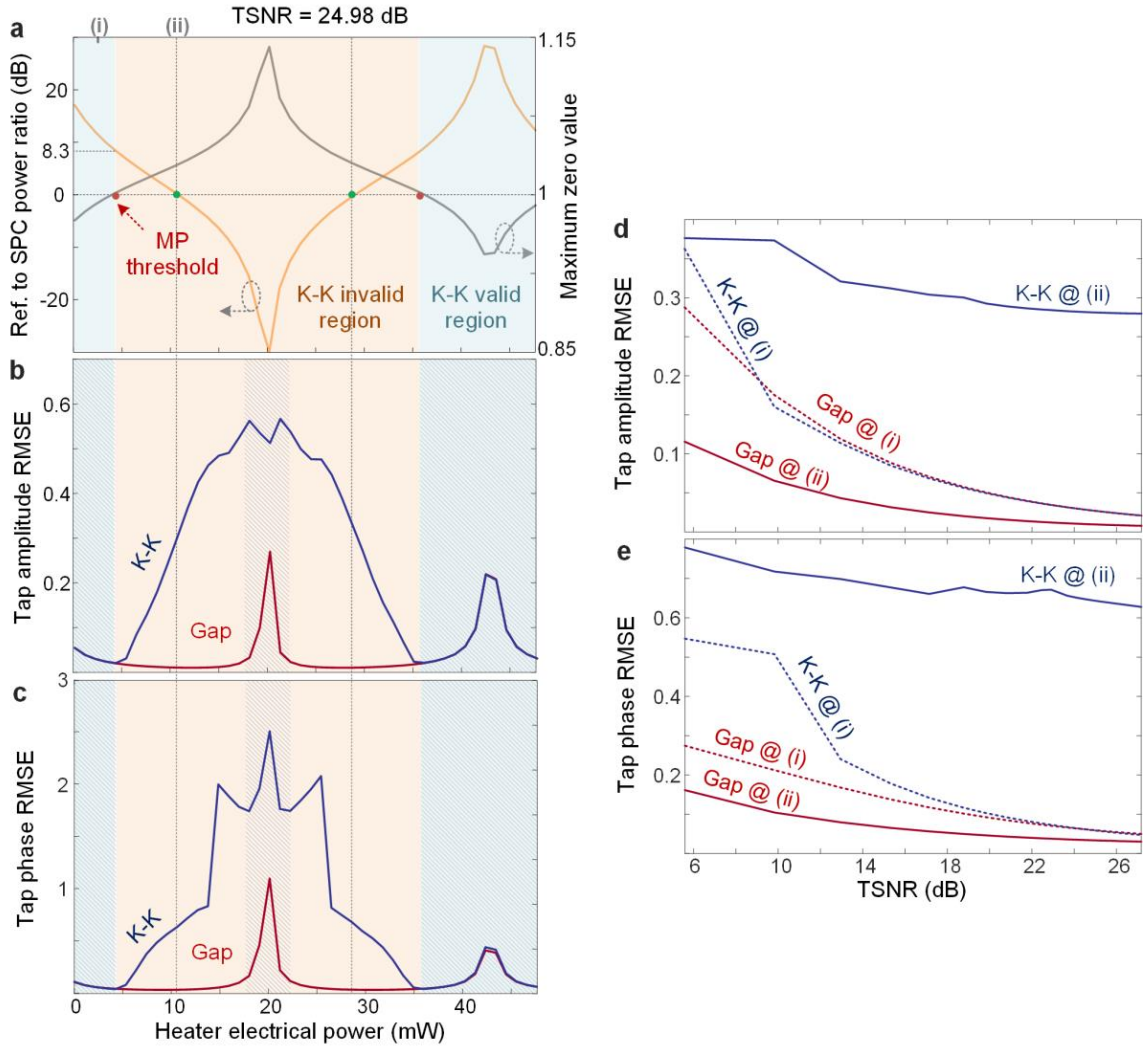

**Figure 3 | Simulated results of PIC characterization using gap and K-K methods.** (a) The simulated power splitting ratio between the reference path and SPC (yellow line) and the maximum value of zeros (grey line). The calculated RMSE of retrieved tap amplitude (b) and phase (c) using the K-K method (blue line) and our gap method (red line), as the power splitting ratio varies. The calculated RMSE of retrieved tap amplitude (d) and phase (e) using the K-K method (blue line) and our gap method (red line), as the *TSNR* of the chip varies, at the heater powers labeled at the top of plot (a); the gap method mostly outperforms the K-K method at (i) and always is best for equal reference and SPC powers (ii).

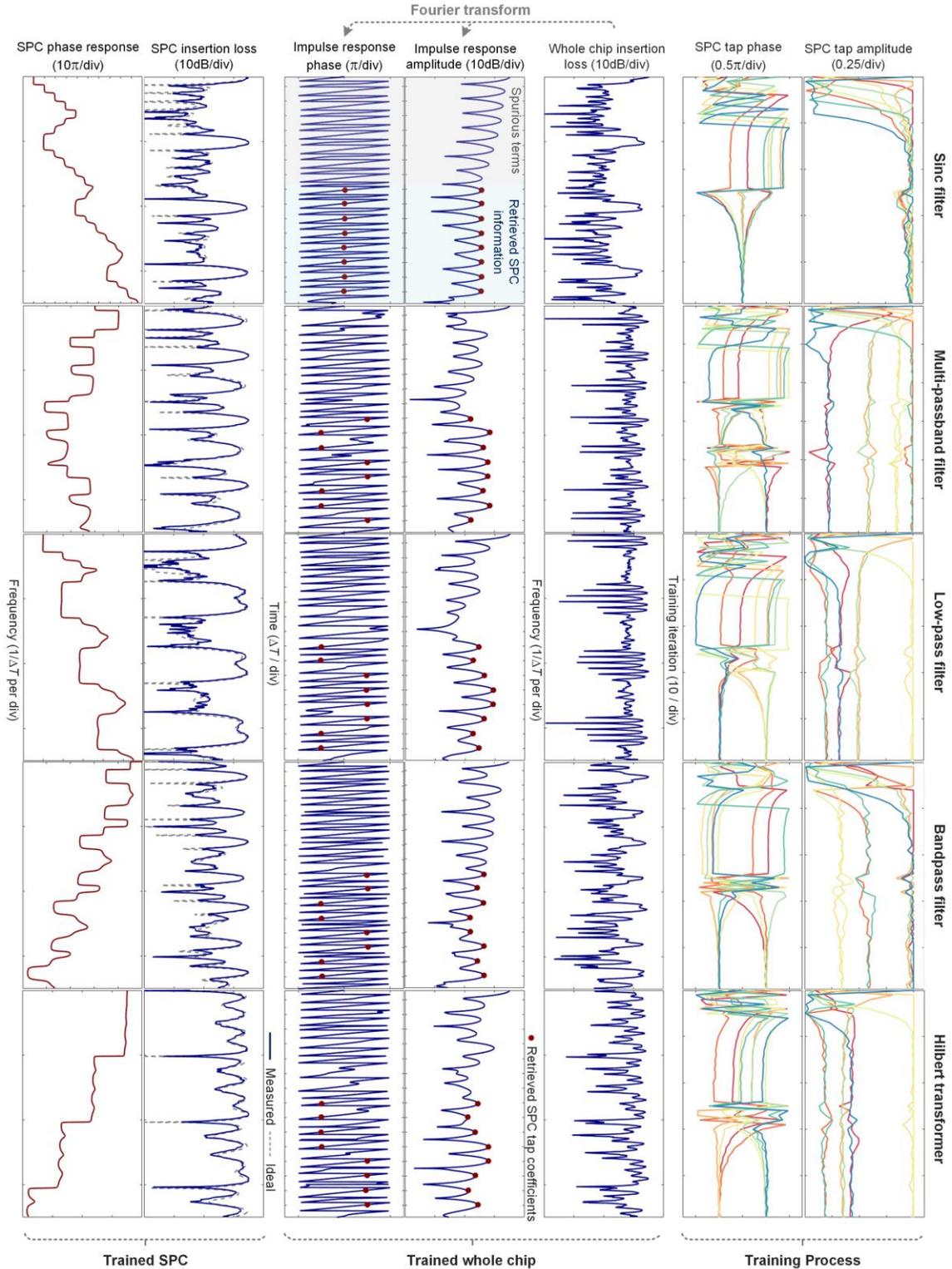

**Figure 4 | Training the PIC using characterized chip information.** The training curve of the SPC taps' amplitudes (first row) and phases (second row). The measured insertion loss spectra of the whole chip (third row), retrieved impulse response amplitude (fourth row) and phase (fifth row), and the SPC's insertion loss spectra (sixth row) and calculated phase response (seventh row), after

the SPC was trained. We note that, during first 25 training iterations, only the MZIs were tuned; during the last 25 training iterations, both the MZIs and phase shifters were tuned.